\input epsf

\magnification= \magstep1  
\tolerance=1600 
\parskip=5pt 
\baselineskip= 6 true mm 

\font\smallrm=cmr8
\font\smallit=cmti8
\font\smallbf=cmbx8

\def\b{\beta}
\def\g{\gamma} \def\G{\Gamma}
\def\d{\delta} 
\def\e{\varepsilon}

\def\l{\lambda} \def\L{\Lambda} 
\def\m{\mu}
\def\f{\phi} 
\def\n{\nu}
\def\j{\psi} 
\def\s{\sigma} 
\def\t{\tau}
\def\W{\Omega}

\def\cl{\centerline}
\def\pa{\partial}

\def\fn#1#2{{\baselineskip=12 pt \footnote{$^ #1 $}{\smallrm #2}}\ }
\def\fnd#1{{\baselineskip=12 pt \footnote{$^\dagger$}{\smallrm #1}}}

\def\half{{\textstyle{1\over2}}}
\def\hhalf{{\scriptstyle{1\over2}}}
\def\qu{\ {\buildrel {\displaystyle ?} \over =}\ }
\def\LL{{\cal L}}
\def\DD{{\cal D}}

{\nopagenumbers

\vglue 1truecm
\rightline{THU-94/15}
\rightline{hep-th/9410038}
\vfil
\cl{\bf GAUGE THEORY AND RENORMALIZATION\fnd{Presented at the 
International Conference on:\hfil\break 
\cl{``The History of Original Ideas and Basic Discoveries 
in Particle Physics"}\hfil\break 
\cl{Erice, Italy, 29 July - 
4 August 1994.}}}

\vskip 1.5 truecm

\cl{Gerard 't Hooft }
\vskip 1 truecm
\cl{Institute for Theoretical Physics}
\cl{University of Utrecht, P.O.Box 80 006}
\cl{3508 TA Utrecht, the Netherlands}
\vfil
\noindent{\bf Abstract}
\smallskip
Early  developments  leading  to 
renormalizable non-Abelian gauge theories for the weak,
electromagnetic and strong interactions, are discussed from a personal
viewpoint. They drastically improved our view of the  role of field
theory, symmetry  and  topology, as well as other branches of
mathematics,  in  the  world  of  elementary particles.

\vfil \noindent{\bf Foreword}
\smallskip
Like most other  presentations  by  scientists  in  this Conference, my 
account of the most important developments that led towards our  present 
view of the fundamental interactions among elementary  particles, is a 
very personal one, recounting discoveries the author was just  about  to 
make when someone else beat him to it.  But there is also something else 
I wish to emphasize. This is the dominant position reoccupied during the 
last 25 years by Theory, in its relation to Experiment. In particular 
Renormalized Quantum Field Theory not only fully regained respectability, 
but has become absolutely essential for  understanding  those basic facts 
now commonly known as ``The Standard Model". I will limit myself only 
to the nicest goodies among the many interesting developments in the theory
of renormalization, and of those I'll only pick the ones that were of 
direct importance to me.

The account given here partly overlaps with a similar expos\'e given two
years ago [1] at SLAC.\vfil\eject}
\pageno=1
\noindent{\bf 1. Renormalization of QED.}
\smallskip
The early days of renormalization theory are, somewhat disrespectfully,
regarded by my generation of physicists as ``prehistory". People
struggled with the phenomenon of the frequent appearance of
``infinities" in their description of relativistic
quantum-electromagnetic interactions. It was natural to attempt to
reformulate the theories such that these nasty infinities disappeared.
A milestone was reached when Hans Bethe [2] found a reasonable looking
expression for the Lamb shift in 1947. 
Julian Schwinger [3] found out how to calculate the first quantum
mechanical corrections sytematically.  Sin-Itiro Tomonaga [4] and
Richard Feynman [5] added a lot to the insights in how renormalization
works.

In those days the only quantum field theory known to be manifestly
physically important was quantum electrodynamics (QED), describing the
interactions between electrons and photons, and a consequence of this
was that many features specific to quantum electrodynamics were thought
to be essential for renormalizability. A technical difficulty was the
problem of ``overlapping divergences", the fact that at higher orders
infinities of different kinds would get entangled. Freeman Dyson [6]
first treated this preoblem in 1949, after which John Ward, Abdus Salam
and Steven Weinberg made improvements. Robert Mills and Chen Ning Yang
in 1966 discovered that these treatments could still develop flaws at
very high order, and they showed how this problem could be handled
(ironically, their paper occurs in overlapping issues of {\it Suppl.
Progr. Theor. Physics} [7]).

An important key in reformulating theories while avoiding infinities 
would later turn out to be the so-called `dispersion relations', 
that had been worked out and discussed by Hans Kramers [8], R. Kronig [9] 
and Nicolaas van Kampen [10].
 
But many physicists became quite unhappy with this course of events.
The resulting framework very much looked as if it could be summarized
as follows:

{\smallskip\narrower\noindent\it
     Start with the ``naive", unrenormalized theory. You  will  see 
     that it contains ``infinities". Renormalization simply amounts 
     to ``subtracting", or ``removing" the infinite terms.\smallskip}

\noindent This sounded like: ``you hit  upon  difficulties;  just
ignore  them, cover them up!". As by miracle,  the  resulting
prescriptions  are  now claimed to be completely unique and
self-consistent. But of  course  the explanations as to why they work
are then lacking,  and  many  textbooks that contain only this
version  of  the  argument  have  added  to  the wide-spread mistrust
and contempt for such an obviously shaky procedure, in
spite of its experimental success, which, according to some,  had  to
be accidental [see the extensive review by Cao and Schweber, ref. 11].

With our present understanding we can resolve most of the conceptual
difficulties people had with renormalization. The complete class of
``renormalizable quantum field theories" is now known, and their unified
treatment gives much more insight. First of all, all these
renormalizable theories should be looked upon as {\it models}, which
have a built-in limitation in that they can often only be treated as a
perturbative series in the coupling constants (with the exception of a
very small subclass, the ``asymptotically free theories", which, unlike
QED, can probably be treated non-perturbatively). The presently adopted
``Standard Model" is just a member of this class describing experimental
observations extremely well, but it is known {\it not} to be infinitely
precise and it will need improvements or even a substitution at
extremely high energies, see further Section 6. Secondly, the problem
of overlapping divergences has now been completely resolved by deriving
dispersion relations directly for the Feynman diagrams, a method not so
well known but extremely important in the consistency proof of these
theories [12].  \bigbreak
\noindent{\bf 2. The Yang-Mills field.}\smallskip\nobreak 

The successes of renormalization theory for QED were so great that
generalizations of this scheme were sought. I presume it was understood
that scalar fields could be added, but they were not observed in QED.
There do exist strongly interacting scalar (or pseudo-scalar)
particles, but strong interactions were still very mysterious. The most
striking feature of QED is gauge-invariance, and this was what led Yang
an Mills to study a theory with a more advanced type of
gauge-invariance, the {\it non-Abelian gauge theory}\fnd{The idea was
preceded 16 years earlier by Oskar Klein, who derived very similar
field equations from a Kaluza-Klein construction [13]. I thank Lev Okun
for pointing this out to me.}.  They wrote what I regard as an
absolutely beautiful paper [14].

Gauge-invariance is generalized the following way. Assume that there
exist more than one type of fermionic fields $\j(x,t)$  (take the
simplest case of just two), which we can arrange as isovectors:
$$\j\ =\ \pmatrix{\j_1\cr \j_2\cr}\ .\eqno(2.1)$$
Consider transformations of the type
$$\j\ \rightarrow\ \W({\bf x},t)\,\j\ ,\eqno(2.2)$$
where $\W$ is a $2\times 2$  (or  possibly larger) matrix. One can then 
construct the covariant derivative $D_\m\j$ as follows:
$$D_\m\j\ =\ (\pa_\m  -ig\,B_\m^a T^a)\,\j\ ,\eqno(2.3)$$

which transforms just as (2.2) if the new fields $B^a_\m$ transform
in  a very special way. Here  $g$  is  just  some  coupling  constant,
and  the matrices  $T^a$   are the generators of infinitesimal
rotations  (2.2).  One can formulate dynamical equations of motion for
the new fields  $B^a_\m$ by first defining the corresponding
generalization of the electromagnetic fields $F_{\m\n}$. Yang and Mills
had first tried
$$ F^a_{\m\n}\ \qu\ \pa_\m B_\n^a \,-\,\pa_\n B^a_\m\ \dots\,,\eqno(2.4)$$
but they quickly ran into problems. The equations one would get were not
gauge-invariant. To their delight however they found that if one adds to
(2.4)
$$+gf^{abc}B^b_\m B^c_\n\ ,\eqno(2.5)$$
where $f^{abc}$ are the structure constants of the Lie group of matrices $\W$ 
in (2.2), gauge invariance is completely restored.

The field equations are generated by the Lagrangian
$$\LL^{\rm
inv}\ =\ -{\textstyle{1\over4}}F^a_{\m\n}F^a_{\m\n}\,-\,{\bar{\j}}(\g_\m
D_\m+m)\j\ ,\eqno(2.6)$$ which is indeed invariant under local gauge
transformations, and as such a  direct generalization of QED.

Since the rigid, space-time independent analog  of  the  transformation 
group (2.2) (henceforth called the  global  group)  was  known  as 
isospin invariance for the strong interactions, Yang  and  Mills  viewed 
their theory as a scheme to turn isospin into a local symmetry, but they 
immediately recognized that then there was  a  problem:  the  Lagrangian 
(2.6)  describes  a  massless  vector  particle  with  three  (or  more) 
components, in general electrically charged as well as neutral ones.  In 
spite of  its  beauty,  this  theory  was  therefore  considered  to  be 
unrealistic. Besides, since these massless particles interact with  each 
other the theory showed horrible infrared divergencies.

When Yang gave a seminar about his recent result in Princeton, one of
the attendants was Wolfgang Pauli. It turned out that Pauli had had
some thoughts about such an approach. Yang recollects [15], as soon as 
he had equation (2.3) on the
blackboard, 

{\smallbreak\narrower\noindent\smallrm  `Pauli asked: ``What
is the mass of this field $\smallrm B_\m$?". I said we did not know.
Then I resumed my presentation, but soon Pauli asked the same question
again. I said something to the effect that that was a very complicated
problem, we had worked on it and had come to no definite conclusions. I
still remember his repartee: ``That is not a sufficient excuse." I was so
taken aback that I decided [...] to sit down. There was general
embarassment. Finally Oppenheimer said, ``We should let Frank proceed".
I then resumed, and Pauli did not ask any more questions.' \smallbreak}

Clearly, Pauli had been studying very similar schemes himself, but had
rejected them because of the mass problem.
Proposals to cure this ``disease" were made several times. 
Feynman, who looked upon this model as a toy  model for quantum 
gravity, proposed to simply add a small mass term  just to avoid the 
infrared problem [16]:
$$\LL\,=\,\LL^{\rm inv}\,-\,\half M^2\big(B^a_\m\big)^2\,.\eqno(2.7)$$

Sheldon L. Glashow [17]  and Martinus J.G. Veltman [18] proposed to use the 
same 
Lagrangian (2.7) as a model for the weak intermediate vector boson. It 
was  hoped  that the mass term  would not  spoil the apparent 
renormalizability of the Lagrangian (2.5). Probably the philosophy  here 
was that the mass term is only a mild symmetry breaking correction of a 
kind we see more often in Nature: isospin invariance itself is also 
softly broken.

Indeed Veltman [17] initially reported progress here: the theory  (2.7) 
is  renormalizable  at  the  one-loop  level.  He  made  use  of   field 
transformations that look like gauge  transformations  even  though  the 
mass term in (2.6) is not gauge invariant:
$${B^a _\m }'\,=\,B^a_\m+gf^{abc}\L^bB^c\m-\pa_\m\L^a\quad;\quad\j'\,
=\,\j+g\L^a T^a\j\,,\eqno(2.8)$$
Here $\L$ may be any function of some arbitrarily chosen field  variable. 
The transformation needed to obtain identities  among  amplitudes for
different  Feynman diagrams (see Fig. 1) were called {\it Bell-Treiman
transformations\/} by Veltman\fnd{using his unique sense of humor.
There never existed any references to either Bell or Treiman.} . It
would have more appropriate if the identities obtained were called {\it
Veltman-Ward identities.}

To me it came as a surprise that  Veltman  managed  to  renormalize
his theory up to one loop with this method. The mass term namely
renders the longitudinal part of the gauge field observable, in spite
of  the  fact that the Lagrangian carries no kinetic term for it. This
theory  should self-destruct. This it does, as Veltman indeed
confirmed, but only if  you try to renormalize diagrams with two or
more loops. To render  the  ``massive Yang-Mills theory" renormalizable,
a better theory was needed.

\midinsert
\epsffile{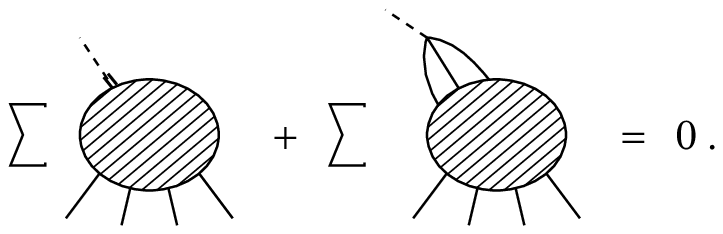}\cl{\smallrm Fig. 1. Veltman-Ward identity among diagrams}
\endinsert

\bigbreak\noindent{\bf 3. The Gell-Mann L\'evy sigma model.}
\smallskip\nobreak
It was one of those caprices of fate that brought me, as a young student 
of Veltman's, to the 1970 Carg\`ese Summer Institute [19].  The champions of 
renormalization were gathered there to discuss the Gell-Mann L\'evy  sigma 
model. This model had been proposed  by  Murray  Gell-Mann  and  Maurice 
L\'evy [20] in 1960. In order to explain the 
existence of a partially conserved axial vector  current  they  added  a 
fourth component to the three pion fields, the sigma field, transforming 
together as a $2\times2$ representation of chiral SU(2)$\times$SU(2).  
The Lagrangian was

$$\eqalign{\LL(\vec\pi,\s,\j,\bar\j) = 
-\half[\pa_\m\vec\pi^2+\pa_\m\s^2]&-\half\m_0^2[\vec\pi^2+\s^2]-
{\textstyle{1\over4}}
\l_0^2[\vec\pi^2+\s^2]^2\cr
-\bar\j[\g_\m\pa_\m+g_0(\s+i\g_5\vec\pi\cdot{\vec\t})]\j&+c\s\,.\cr}
\eqno(3.1)$$

If we take $\m_0^2$ here negative then the potential for the scalar fields 
has the by now familiar dumb-bell shape. The sigma field gets a vacuum 
expectation value, $$\langle\s\rangle\,=\,F\,=\,|\m_0|/\l_0\,,\eqno(3.2)$$
so in a perturbative expansion we write $\s=F+s$, and expand in $s$. 
The nucleon fields $\j$ get a mass $g_0F$,  the  pions  have  a  tiny 
mass-square proportional to the small constant $c$, whereas the sigma 
field $s$ becomes a heavy resonance.

Benjamin W. Lee [21], Jean-Loup Gervais [22] and Kurt Symanzik [23] 
explained 
in their Carg\`ese lectures how this model could be renormalized, and that 
its beautiful features would not be seriously affected by 
renormalization. It was clear to me at that time that one can produce 
mass terms for Yang-Mills fields in a way very similar to this sigma 
model. I did not ask many questions in this School, but I did ask one 
question to Benjamin Lee and to Kurt Symanzik:  ``Do  your  methods  also 
apply to the Yang-Mills case?" They both gave me the  same  answer:  ``If 
you are Veltman's student, you should ask him; I am  not an expert in 
Yang-Mills theory."

\bigbreak\noindent{\bf 4. Massless Yang-Mills.}\smallskip

     This I did, as soon as I was back in Utrecht. Now  Veltman  was
skeptical about spontaneous symmetry breakdown in particle 
theory. His opinion was that if that happens the vacuum would have a 
tremendously large energy density, and  this would give the physical 
vacuum an enormously large cosmological coupling constant. 

But we know it happens in the sigma model, which  describes strong 
interactions pretty nicely. And if it is not symmetry breaking  then  at 
least all other  vacuum  fluctuation  effects  also  contribute  to  the 
cosmological coupling constant, not as much as in a weak interaction 
theory with Higgs mechanism, but still far more  than the experimental 
upper bound. The cosmological coupling constant problem should be 
postponed until we solve quantum gravity; we should not let it affect 
our theories at the GeV or TeV scale.

Veltman still had some reservations concerning theories with
spontaneous symmetry breakdown coupled to Yang-Mills fields. He
told me he had now nearly convinced himself that you can't add
scalar particles to renormalize massive Yang-Mills fields unless you
gave them the wrong metric, which would be unacceptable. Would the
Higgs theories be the recipe to avoid such dangers? Clearly, I needed to understand these systems better myself. It was then
decided what my research program would be. First I would try to
really understand all details of the massless, unbroken Yang-Mills
system, for which Veltman gave me his blessing, and then I would add the 
mass, by a ``spontaneous local symmetry breaking mechanism".

The status of pure Yang-Mills theory was somewhat vague. Strong formal
arguments existed that this ``theory"\fn*{Unfortunately in modern
scientific papers the meanings of the words ``model" and ``theory" are
being interchanged for reasons with which I do not agree but that can
be explained psychologically.} had to be renormalizable.  But there
were competing and conflicting theories as to what its Feynman rules
were. One paper on this subject was a short Physics Letters paper by
Ludwig D. Faddeev and Victor N. Popov [24].  It was all I needed to
understand what was going on. Faddeev and Popov argued that a gauge
invariant functional integral expression for the amplitudes had to have
the form
$$\G\,=\,\int e^{\,i\int\LL^{\rm inv}(A){\rm d}^4x}\prod_x{\rm d}^\ell 
A(x)
\,,\eqno(4.1)$$
where $A(x)$ stands for the $\ell$ field components of the gauge 
and matter 
system. However, since the integrand is invariant under gauge 
transformations, one only needs to integrate over the inequivalent field 
configurations, each being constrained by some gauge condition. As a 
gauge condition one typically takes
$$\pa_\m B^a_\m\,=\,0\,.\eqno(4.2)$$
If we impose this constraint on  the integrand one needs a Jacobian 
factor however. So if we keep track of the  measure, this turns  the 
integral into
$$\G\,=\,C\int e^{\,i\int\LL^{\rm inv}(A){\rm d^4x}}\prod_x 
{\rm d}^\ell A(x)
\d\big(\pa_\m B^a_\m(x)\big) {\rm det}\left({\pa(\pa_\m B^a_\m)\over 
\pa\L}\right)
\,.\eqno(4.3)$$ The theory produces a transverse propagator:
$${\d_{\m\n}-{\displaystyle{k_\m k_\n\over k^2-i\e}}\over k^2-i\e}\,.
\eqno(4.4)$$
Other theories [25]  led to a Feynman gauge propagator,
$${\d_{\m\n}\over k^2-i\e}\,,\eqno(4.5)$$
and how this could be related to a functional integral was not clear. 
More important, I thought, was that none of the existing papers 
provided 
for a precise  prescription  as to  how the infinities should be 
subtracted. The formal arguments were there, but how does it work in 
practice?

This became the subject of my first publication [26]\fn\ddagger{In this, and
several of my subsequent publications, I benefitted enormously from
Veltman's intensive ineterest and advices.}. The answer to this
question was indeed far from trivial. Several  things had to be done.
First, the formalism to obtain the  Feynman rules from the functional
integrals could be simplified. The existing procedure to deduce the
ghost Feynman rules from the determinant was not satisfactory. I
observed that one can write  $$\big({\rm det}\,{\cal M}\big)^{-N}\,=\,
C\int \DD\vec\f\DD{\vec\f}^*
e^{-{\vec\f}^*{\cal M}\vec\f}\,,\eqno(4.6)$$
where $\f$ is a complex Lorentz-scalar field with $N$ components 
($\DD\f$ stands for the functional measure $\prod_x{\rm d}\f$). One now
reads off directly the Feynman rules for closed loops of $\f$ fields. A
factor $N$ goes with each closed loop.  Since we want $N$ to be $-1$
our closed loops will usually go with a factor $-1$, just like the
rules for fermions. Indeed, one can also write
$${\rm det}\,{\cal M}\,=\,C\int\DD\eta\DD\bar\eta\,e^{\bar\eta{\cal
 M}\eta}
\,,\eqno(4.7)$$
where $\eta$ is an anticommuting (Grassmann) variable.

Next, I could also see how Faddeev and Popov's trick could  produce 
the Feynman gauge. Just take auxiliary field variables $F^a$ and impose 
the gauge  
$$\pa_\m B^a_\m\,=\,F^a\ .\eqno(4.8)$$
One then sees that
$$e^{\hhalf (\pa_\m B_\m^a)^2}\,=\,\int\DD F e^{-\hhalf F^2}
\d(\pa_\m B_\m^a-F^a)
\,.\eqno(4.9)$$

To see that the renormalization counter terms do  not spoil gauge 
invariance we needed Ward identities. It turned out to be sufficient to 
prove identities of the form of Fig. 2.

\midinsert
\epsffile{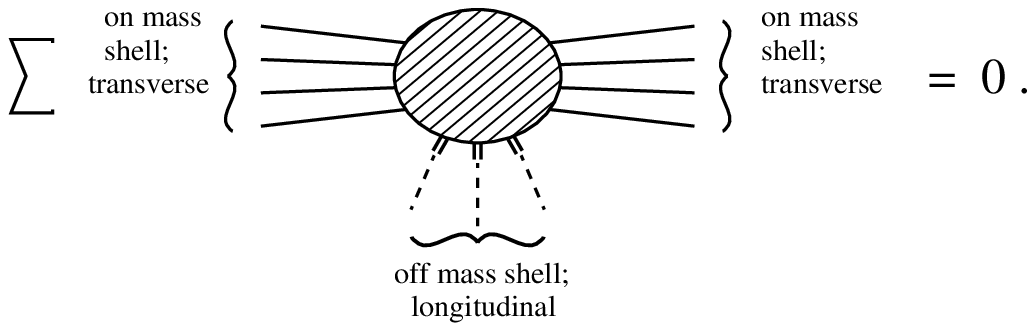}\cl{\smallrm Fig. 2. The Ward identity for Abelian and 
non-Abelian gauge theories.} \endinsert

Since between the accolades reducible and irreducible diagrams must  all 
be added together, these  identities  are  sufficient to restrict all 
counter terms completely up to gauge invariant ones. This point was 
often not realized by later investigators. It becomes obvious if we
 expand
the diagrams of Fig. 2 in terms of one-particle-irreducible ones.

The proof of these Ward identities was much more complicated than 
the Veltman-Ward identities mentioned  before, because we had to 
disentangle carefully the contributions of various ghost lines. See 
Fig. 3, which was an intermediate step.

\midinsert
\epsffile{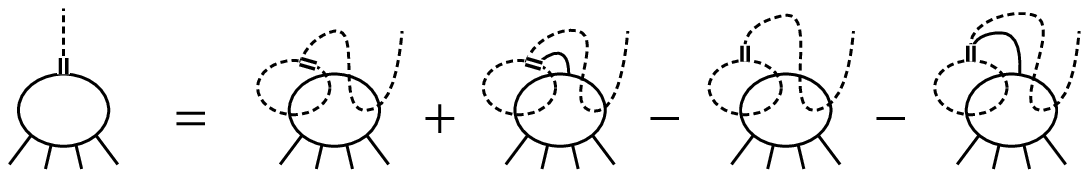}\cl{\smallrm Fig. 3. Part of the proof for Fig.2.}
\endinsert

I was annoyed that I couldn't use a simple symmetry argument for the 
proof as Veltman had done for his case. Only much later it  was 
discovered how to do this. C.~Becchi, A.~Rouet and Raymond Stora found that
the underlying symmetry for this identity is an anticommuting one. Their 
marvelous discovery was this [27]. Take as an invariant Lagrangian for 
instance
$$\LL^{\rm inv}\,=\,-{\textstyle{1\over4}}F^a_{\m\n}F^a_{\m\n}-D_\m\f^*D_\m\f
-V(\f,\f^*)-\bar\j(\g D+m)\j +\dots\ ,\eqno(4.10)$$
and add as a gauge fixing term 
$$\LL^{\rm gauge}\,=\,-\half(\ell^a)^2\,,\eqno(4.11)$$
where $\ell^a$ is anything like $\pa_\m B_\m^a\ $, $\ B_4^a\ $, etc. 
Introduce the  
ghost fields $\eta$ and $\bar\eta$ which must be anticommuting.  
Consider then the anticommuting variation:
$$\eqalignno{\d B_\m^a&=\,\bar\e D_\m\eta^a\,;&(4.12a)\cr
\d\f&=\,-ig\bar\e T^a\eta^a\f\,;&(4.12b)\cr
\d\eta^a&=\,\half g\bar\e f^{abc}\eta^b\eta^c\,;&(4.12c)\cr
\d\bar\eta^a&=\,\bar\e \ell ^a(B,\f,\dots)\,.&(4.12d)\cr}$$
Here the first lines, eqs. (4.12a) and (4.12b), are just gauge 
transformations;
$\bar\e$ is an infinitesimal, anticommuting symmetry generator. 
Then  the total Lagrangian of the theory when taken to be
$$\LL\,=\,\LL^{\rm inv}+\LL^{\rm gauge}+\LL^{\rm ghost}\,,\eqno(4.13)$$
with
$$\LL^{\rm ghost}\,=\,-\bar\eta^a\d\ell ^a(B,\f,\dots\eta)\,,\eqno(4.14)$$
is invariant under this {\it global} transformation. The above 
identities are 
nothing but an expression of this invariance, now called BRS invariance.

I had to convince myself that the rules obtained produced a unitary 
theory. The new identities were sufficient to guarantee this.  Just  one 
problem remained:  the identities overdetermined  the renormalization 
counter terms. Would there never be a conflict? There was a well-known 
example of just such a conflict in the literature: the Adler-Bell-Jackiw 
anomaly. Steve Adler [28], and independently from him John Bell 
and  Roman 
Jackiw [29], had discovered that diagrams of the kind depicted in Fig. 4 
cannot be renormalized in such a way that both the vector current and 
the axial vector current are conserved. If  something like this would 
happen in a gauge theory there would be deep trouble. I could prove that 
if no gauge fields are coupled to the axial charge, clashes of this sort 
will not destroy renormalizability in diagrams with up to one loop. The 
trick was to use a fifth dimension for the internal lines inside the 
loop.

\midinsert
\epsffile{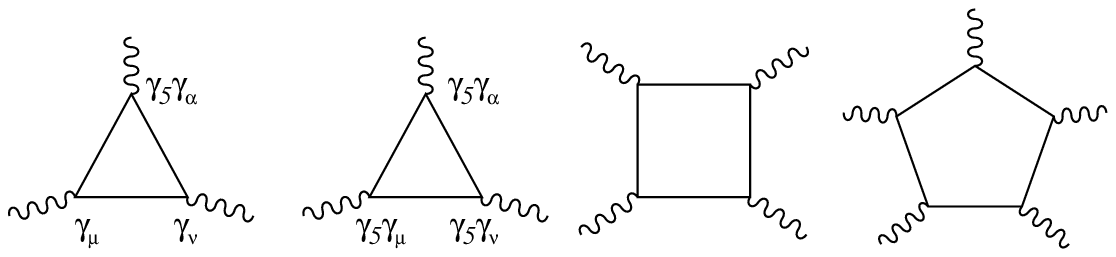}\cl{\smallrm Fig. 4. Diagrams with ABJ anomalies.}
\endinsert

What if you have more than one loop? I tried to use six, seven or 
more dimensions but this does not work . I was confident that the problem 
could be solved, but was unable to do it then.

Soon after my paper had come out, two other papers appeared, one by 
Andrei A. Slavnov [30]  and one by John C. Taylor [31] . Both 
observed that the 
identities I had written down could be generalized.  If some of the 
external lines are neither longitudinal nor on mass shell one gets extra 
contributions where the ghost line ends up at one of  these  lines. See Fig. 5.

\midinsert
\epsffile{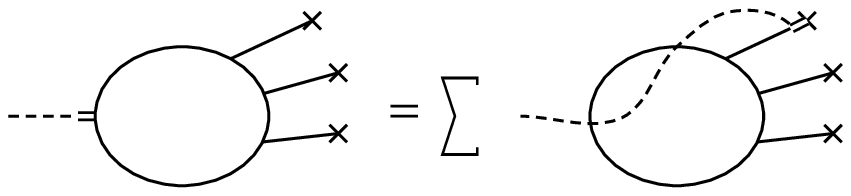}{\smallrm \cl {Fig. 5. Slavnov-Taylor identities.}\break
\cl{ The summation is over the various sources to which the dotted line 
can be attatched.}}
\endinsert

The derivation went the same way as that for my own identities. The only 
reason why I had not written these identities in this new form before 
was that I thought the extra pieces would be cumbersome,  requiring  new 
renormalization counterterms of their own, and,  furthermore,  I  didn't 
need them. I think I was still suffering from the indoctrination that
infinities should be avoided at all costs. It is clear now that these newer  
identities  are  more complete. And so it happened that they  were to 
become  known  as the Slavnov-Taylor identities, a pivotal property of
any gauge theory.
\bigbreak

{\bf 5. The Higgs-Kibble Mechanism and Dimensional Renormalization.}
\smallskip
For  my  advisor, Veltman, all this was just Spielerei. Massless 
Yang-Mills fields seem not to occur in Nature. They are just there for 
exercises. The real thing is the massive case, and he thought that that would 
be an entirely different piece of cake.  Actually however, the step 
remaining to be taken was a small one [32]. As I  knew  from  Carg\`ese, the 
actual nature of the vacuum state has little effect upon renormalization 
counter terms. All needed to be done is to add to the gauge invariant 
Lagrangian the by now familiar Higgs terms,
$$\LL^{\rm Higgs}\,=\,-\half(D_\m\f)^2-V(\f)\,,\eqno(5.1)$$
where $V(\f)$  has the familiar dumb-bell shape, just like in the 
Gell-Mann L\'evy sigma model (for simplicity I take the $\f$ field 
here to be 
a real multiplet). Writing $\f = F + \eta$ we  get a gauge invariant 
Lagrangian for the $B$ and $\eta$ fields, such that now the $B$ fields get 
the required mass.  To appease Veltman I wrote the self-interaction as
$$V\,=\,{\textstyle{1\over8}}\l(2F\eta+\eta^2)^2\,,\eqno(5.2)$$
so that at least at lowest order the vacuum energy density vanishes.  In 
terms of the $\eta$ fields the gauge transformation laws for these and the  
$B$ field look very similar:
$$\eqalign{{B_\m^a}'&=\,B_\m^a+f^{abc}\L^bB_\m^c-{1\over g}\pa_\m\L^a\,;\cr
\eta'&=\,\eta+T^a\L^a\eta+T^a\L^aF\,.\cr}\eqno(5.3)$$
Everything else went exactly as in the previous paper. Because the local 
gauge invariance is still exact we again have Slavnov-Taylor identities, 
BRS invariance and from them one can prove unitarity and equivalence of 
the various gauge choices. A judicious gauge choice was found such that 
the propagators for the massive gauge fields and the other fields became 
as simple as possible.

The problem of regularizing and renormalizing diagrams with two  or 
more loops was still there. Veltman and I discussed  a lot about this 
problem and eventually agreed what the best strategy was: continuous 
variation of the number of space-time dimensions [33]. Upon Veltman's 
explicit
instructions, both verbally and in writing [34], I refrain from commenting 
about who of us did what in that paper (although one doesn't forget that
easily, as he claims). 

As if it were a seed from outer space, the idea of making the 
dimensionality of space-time continuous germed simultaneously in various 
places as an  answer to different problems.  Kenneth G. Wilson
and Michael E. Fischer [35] were  writing  a  paper  proposing  to
calculate  critical phenomena in statistical physics in $4-\e$
dimensions as an expansion in $\e$. And independently of us C.~Bollini  
and  J.~Giambiagi [36],  and  J.~Ashmore [37], also suggested to use
analyticity in space-time dimensions as a regulator.

Dimensional regularization does {\it not} work if there is an
 Adler-Bell-Jackiw
anomaly. Initially I thought that this could not be more than a technical
difficulty, but it was quickly pointed out that indeed if these anomalies
do not cancel gauge theories cannot be renormalized at all [38]; they 
are sick.
Consequently in all gauge theories one must impose the new and important 
requirement that these chiral anomalies cancel out against each other. Only
at later times the physical reasons for this requirement became evident: it
has everything to do with the appearance of instantons in the theory. The
symmetries destroyed by them cannot be gauged.

It may be noticed that by now I entirely  address  the  problem  of 
renormalization as a procedure for infinity subtraction. As explained in 
the beginning this is not at all what renormalization really is  from  a 
physical point of view. It is preferable to talk about regularization 
first, and then renormalization afterwards. Regularization is the 
replacement of a theory by a slightly mutilated model, using a cut-off. 
In such a model the observed parameters are finite and related in some
calculable way to the ``bare" parameters. 

We must show that, even if the relations between the observed and the
bare parameters tend to become divergent, the effects of the cut-off
become negligible at  large distance scales. One only needs to
demonstrate that in terms of the observed, ``renormalized" observables
the  limit  where the cut-off goes away exists and is perturbatively
finite\fnd{{\smallit non-}perturbatively the limit might not exist. This
happens for instance in $\l\f^4$ theory. Such a theory is perturbatively 
fine
but non-perturbatively non-consistent. Curiously, such models are 
still extremely
useful in physics; their predictions are accurate but cannot be infinitely
accurate, see sect. 6.}. It  does  not matter much how crazy the 
mutilation was in the
beginning, as long as the limit is well-behaved. Going to $4-\e$
dimensions is just such a crazy regularization scheme. It turns out
to  be  technically  extremely elegant. Anyway, the important thing was
that this method works fine  at all orders of perturbation expansion
and not just up to one loop, like the 5 dimensional procedure found
earlier.

We now had a general scheme for producing theories with interacting 
massive vector particles  . At first I was thinking about applying it to 
rho mesons, as a nice generalization of the Gell-Mann L\'evy sigma  model. 
But of course Veltman could convince me that the weak interactions  were 
a  much  more  promising  application.  I had essentially reproduced 
Weinberg's model before I saw his 1967 paper [39]. Veltman brought it with
him when he returned from Geneva. Anyway, when my paper on the Massive
Yang-Mills fields came out I soon received letters from both Weinberg
and Salam [40] with copies of their 1967 papers, in which the 
renormalizability
of weak interaction theories of this nature had been speculated on.

More important to my mind was that we now  had a large class of 
renormalizable models with  massive and massless vector mesons. A 
crucial argument was added to this by Chris Llewellyn Smith [41] and  J. 
Cornwall, D. Levin and George Tiktopoulos [42] : they showed that requiring 
unitarity implies that the {\it only} such models are gauge theories. So not 
only do we have a large class of new models, we have the {\it complete} class 
of renormalizable vector theories.
\bigbreak\noindent
{\bf 6. The Renormalizability Requirement.}
\smallskip

Having been so successful in formulating the renormalizability requirement
and introducing this in realistic theories for the weak interaction it
would have been natural to promote renormalizability to the status 
of a primary 
principle in quantum field theory. Requiring such a theory to be 
renormalizable 
indeed limits us to a subclass large enough to enable us to include 
all particle
and interaction types known, and small enough to be fantastically predictive.
The extremely accurate tests of the Standard Model at facilities such as
LEP confirm the basic correctness of such a standpoint.

Nevertheless such a dogmatic attitude would obscure some important aspects
of quantum field theories in general. It cannot be ignored that even
renormalized theories suffer from weaknesses undermining their status as
ultimate descriptions of reality. What I am referring to here is the fact 
that these theories can be rigorously formulated as perturbative expansions
in terms of small coupling parameters, but most of them will probably not
allow any meaningful definition at all beyond this perturbative level. 
There
is little harm in this in practice because the margins within which these
theories are ill defined will be of order $e^{-C/g^2}$ where $g$ is 
the coupling
constant and $C$ some characteristic number related to the theory's 
$\b$-functions.

It is safe to say that at present there exists no mathematically 
`perfect' quantum field theory at all in four space-time dimensions 
(unless it were to describe only free particles). At least five degrees of 
sophistication can be distinguished:\smallskip
\item{1.} {\it Non-renormalizable} theories. These are theories for which
the amplitudes can only be computed at tree level:
$$ \G\,=\,a_1g^2+{\cal O}(g^4).$$ The higher order quantum corrections of 
order $g^4$ would require new and unknown subtraction constants. 
Examples are:
the old Fermi weak interaction theory in hich four fermions are coupled at 
single
points, and gravitation coupled to matter. Note that even these theories,
in spite of their limitations, have proven their worths as stages of 
understanding
in physics. But the region of applicability ends at energy scales of order
of $1/g$ times the typical mass scale.
\item{2.} {\it One-loop renormalizable} theories:
$$ \G\,=\,a_1g^2+a_2g^4+{\cal O}(g^6).$$ No new counter terms are 
needed even at 
the one loop level. Examples are pure gravity and massive Yang-Mills 
fields with the Higgs particle omitted.
\item{3.} {\it Renormalizable} theories. No new counter terms at any
finite order. Examples are QED, $\l\f^4$ theory and
all gauge theories. The asymptotic expansion is well defined:
$$ \G\,=\,\sum_na_ng^{2n} + {\cal O}\big(e^{-C/g^2}\big)\,,$$
but the marginal non-perturbative correction terms are still beyond control.
\item{4.} {\it Asymptotically free} theories [43]. In these theories all 
running
coupling constants tend to zero as the energies go to infinity. In these
theories the renormalization counter terms can be rigorously defined in
the limit of a vanishing cut-off, and in view of this one may well suspect 
that these theories can be definied rigorously mathematically. Unfortunately
there is no proof of that\fnd{The difficulty one encounters when trying
to prove
mathematical completeness of an asymptotically free theory reside in the
{\smallit a priori} unknown characteristics of the vacuum state}. Indeed, 
such theories 
are usually not Borel resummable. Examples are: QCD and some special 
renormalizable 
models containing not only gauge fields but also scalars and spinors.
\item{5.} {\it Borel resummable} theories. In these theories one redefines
the amplitudes $\G$ as
$$\G\,=\,\int_0^\infty{\rm d}z\,e^{-z/g^2}B(z)\,,$$
where $B(z)$ has a convergent expansion in powers of $z$, but it has to be 
proven that it is well-defined for all $z$. There are no known non-trivial
examples in four dimensions, but an asymptotically free model in the
planar limit ($N\rightarrow\infty$) behaves this way [54].
\smallskip

\midinsert
\epsffile{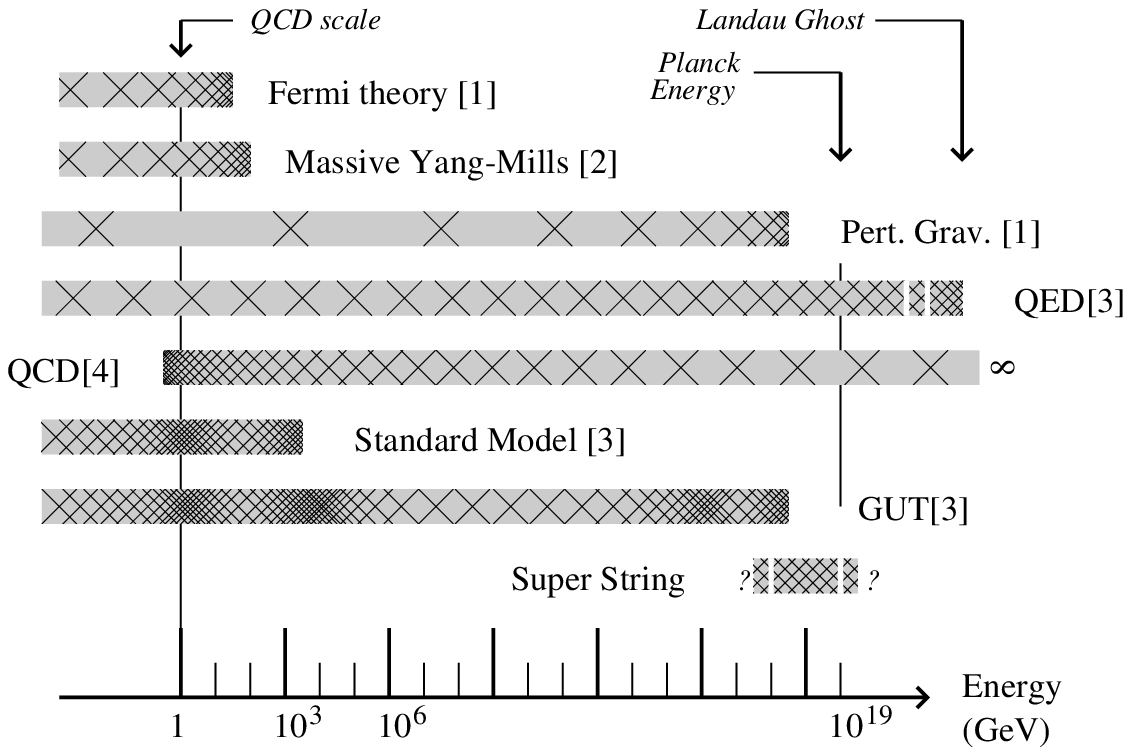}
\narrower\noindent{\smallrm Fig. 6. Comparison of the domains of validity 
for different 
quantum field theories. Shading indicates amount of 
structure in a model. Numbers
correspond to the position in the list of increasing sophistication.}
\endinsert

\noindent A theory is more sophisticated if it allows for more structure
in a larger domain of different energy scales. If the structure becomes
too complex the mathematical rigour breaks down because of lack of 
convergence
of the known expansion techniques for their analysis. A theory 
further down
the list I just gave allows for more structure along a larger 
domain of energies.
See Fig. 6 in which the coloration indicates structure.

In summary: the more complex the particle spectrum and interactions are, and
the more rigid our demands for precision, the stronger the constraints on a
theory. Renormalizability is not sufficient to make a theory completely
airtight but it provides us with a paradigm that allows consistent 
calculations
with margins that decrease exponentially as the interaction 
strengths decrease.
All our theories have a built-in twilight zone, where, according 
to their own
logic, the calculational rules fail. In these regions our theories 
will have to be
supplanted by others, as has always been the case in physics. It is 
suspected,
but has never been proven, that asymptotically free theories have no such 
twilight zone.
If one merely has a renormalizable theory, the twilight zone is 
exponentially far away.
\bigbreak\noindent
{\bf 7. Further developments}
\smallskip

In practice this new development was an enormous improvement.
 Within just a 
couple of years many further insights were obtained. A charming feature of
many versions of `unified gauge theories' is that they allow for entirely new
kinds of solutions for their field equations corresponding to particles
with a single magnetic (north or south) charge: magnetic monopoles [44].
Independently of the present author the discovery was made when Alexander A.
Polyakov introduced his ``hedgehog solutions'' [45]. As he states in a 
footnote
of his paper, the fact that his solution possesses magnetic charge was
pointed out to him by Lev Okun'.

Further search brought to light that there exist also four-dimensional
localized field solutions [46] in all non-Abelian gauge theories. When I
introduced the name `instantons' for these the word was censored by
Phys. Rev. Letters [47]. They wanted to change it in: ``Euclidean Gauge
Pseudoparticle Solution, EGPS for short". When I protested they went as far
as ``Euclidean Gauge Soliton (EGS)''. These ugly acronyms 
only hastened the general acceptance of `instanton' ever since.
My interest in the instanton, and its peculiar effects on fermions, 
came from the
fact that these effects are exponential in $-1/g^2$. How can this be 
squared with
the observation of the previous chapter, which was that effects this small
need not be calculable at all in a renormalizable field theory? I
decided to go through a complete calculation to check whether 
there would be any
residual infinities or other ambiguities [48]. There were none. Just
because instanton effects either violate a symmetry such as baryon
number conservation or lift a degeneracy such as in pure QCD, the effects
turn out to be totally unambiguous.

It is a characteristic of successful theories that they provide  further 
understanding in many different areas  of  the  field,  in  elegant  and 
unsuspected ways. As for the Standard Model, we now know that the  roles 
of asymptotic freedom, monopoles  and  instantons  are  crucial  in  our 
present picture of quark confinement, the hadron spectrum,  the  scaling 
phenomena  and  jet  physics.  The  renormalized  theory  allows  us  to 
reproduce the observed data on the  Z  and  W  bosons with unprecedented 
precision. The Standard Model, as a gauge theory with  fermions  and  at 
most only one scalar, is indeed tremendously successful.
     My  presentation  was  sketchy.  Subjects  that, among others, should  
have  been discussed as well or in more detail, but for which I had no time 
left, are:\smallskip\noindent
$\bullet\ $   The discovery of asymptotic freedom [43, 49];
\hfil\break  $\bullet\ $   Renormalization of quantum gravity [50];
\hfil\break  $\bullet\ $   Quark confinement [51];
\hfil\break  $\bullet\ $   Constructive Field Theory [52];
\hfil\break  $\bullet\ $   Appelquist-Carrazone decoupling [53];
\hfil\break  $\bullet\ $   The $N\rightarrow\infty$ limit and the 
resulting planar 
diagram approach  to QCD [54]; 
\hfil\break  $\bullet\ $   Naturalness and the instanton index 
matching condition [55];
\hfil\break  $\bullet\ $   $B+L$ violation [56] and monopole catalysis [57];
\hfil\break  $\bullet\ $   Renormalons [58]. 
\smallskip
All these topics are closely linked to the renormalization issue.
Of course, after two decades have past, also  the  deficiencies  in 
our theories are standing out clearly: a theory that  should  explain  why 
the local symmetry is as it is, where the fermion  spectrum  comes  from 
and how the values of about 20 constants of nature  are  determined,  is 
still being searched for, but it is difficult to believe  that  a  giant 
leap in particle theory, as it occurred in the 70's, will be repeated in 
the near future.
\bigbreak\noindent
{\bf References}\smallskip\smallrm
\baselineskip=12 pt 
\item{1.} G.~'t~Hooft, "Renormalization of Gauge Theories", in
Proceedings of the Third International Symposium on the History of
Particle Physics: 'The Rise of the Standard Model", SLAC, June 24-27,
1992.
\item{2.} H.A. Bethe, {\smallit Phys. Rev.} {\smallbf 72} (1947) 339.
\item{3.} J. Schwinger, {\smallit Phys. Rev.} {\smallbf 73} (1948) 416, 
{\smallit ibid.} {\smallbf 74} (1948) 1439.
\item{4.} S.~Tomonaga, {\smallit Progr. Theor. Phys.} {\smallbf 1} (1946) 27.
\item{5.} R.P.Feynman, {\smallit Rev. mod. Phys.} {bf 20} (1948) 367; 
id, {\smallit  Phys. Rev.}  {\smallbf 74} (1948) 939, 1430.
\item{6.} F.J.~Dyson, {\smallit Phys. Rev.} {\smallbf 85} (1952) 631.
\item{7.} R.L.~Mills and C.N.~Yang, {\smallit Suppl. Progr. Theor. Phys.} 
{\smallbf 37} and {\smallbf
38} (1966)  507.
\item{8.} H.~Kramers, {\smallit Quanten theorie des Elektrons und der
Strahlung}, Akad.Verlag, Leipzig (transl. D.~ter Haar, North Holland,
Amsterdam, 1957);  id, A review talk at the Shelter Island conference
(June 1947), unpubl. See S.S.~Schweber, "A short history of Shelter
Island I", in {\smallit Shelter Island II} (eds. R.~Jackiw, N.N.~Khuri, 
S.~Weinberg and E.~Witten, MIT Press, Cambridge, Mass, 1985). See also his
{\smallit Collected scientific papers} (Amsterdam 1956) 333 and 347.
\item{9.} R.~de L.~Kronig, {\smallit Journ. Opt. Soc. Amer.} {\smallbf 
12} (1926)  547.
\item{10.} N.G.~van Kampen, {\smallit Ned. T. Natuurk.} {\smallbf 24}, 
Januari 1958 and 
Februari 1958 (in Dutch).
\item{11.} T.Y.~Cao and S.S.~Schweber, "The Conceptual Foundation and
Philosophical Aspects of Renormalization Theory", Brandeis and Harvard 
Universities preprint.
\item{12.} M.~Veltman, {\smallit Physica} {\smallbf  29} (1963) 186; 
G.~'t~Hooft 
and M.~Veltman,  
"DIAGRAMMAR",  CERN  Report  73/9  (1973), 
     reprinted in "Particle Interactions at  Very  High  Energies",  Nato 
     Adv. Study Inst. Series, Sect. B, vol. {\smallbf 4b}, p. 177; 
G.~'t~Hooft, "Gauge  Field  Theory",  in  {\smallit Proceedings  of  the  
Adriatic 
     Meeting}, Rovinj 1973, ed. M.~Martinis et al, North  Holland  /  Am. 
     Elsevier, p.321; 
G.~'t~Hooft,  "DIAGRAMMAR  and  Dimensional  Renormalization, in 
"Renormalization and  Invariance  in 
     Quantum Field Theory", Capri Summer Meeting  July  1973,  ed.  
E.R.~Caianiello, Plenum New York 1974, p. 247.
\item{13.} O. Klein, in ``New Theories in Physics", 
Conference organised in collaboration 
with the International Union of Physics and the Polish Intellectual 
Co-operation 
Committee, Warsaw, May 30th - June 3rd, 1938.
\item{14.} C.N.~Yang and R.L.~Mills, {\smallit Phys. Rev.} {\smallbf 96} 
(1954) 191,  see
also R.~Shaw,  Cambridge  Ph.D. Thesis, unpublished.
\item{15.} C.N.~Yang, Selected Papers 1945-1980 With Commentary, 
Freeman and Co., San
Francisco 1983, p. 20.
\item{16.} R.P.~Feynman, {\smallit Acta Phys. Polonica} {\smallbf 24}
 (1963) 697.
\item{17.}  S.L.~Glashow, {\smallit Nucl. Phys.} {\smallbf 22} (1961) 579.
\item{18.}  M.~Veltman, {\smallit Nucl. Phys.} {\smallbf B7} (1968)
637; J.~Reiff  and  M.~Veltman, {\smallit Nucl. Phys.} {\smallbf B13}
(1969) 545; M.~Veltman, {\smallit Nucl. Phys.} {\smallbf B21} (1970) 288.
\item{19.}  Proceedings of the Carg\`ese Lectures in Physics, Vol. 5,
Gordon and Breach, New York, London, Paris, 1972, ed.~D.Bessis.
\item{20.}  M.~Gell-Mann and M.~L\'evy, {\smallit Nuovo Cim.} {\smallbf 16} 
(1960) 705.
\item{21.}  B.W.~Lee, {\smallit Nucl. Phys.} {\smallbf B9} (1969) 649;
B.W.~Lee, ``Chiral Dynamics", Gordon and Breach, New York, London,
Paris 1972.
\item{22.}   J.-L.~Gervais and B.W.~Lee,  {\smallit Nucl. Phys.} 
{\smallbf B12} (1969) 627.
\item{23.}   K.~Symanzik, {\smallit Let. Nuovo  Cim.} {\smallbf 2} 
 (1969)  10,  id,
     {\smallit Commun. Math. Phys.} {\smallbf 16} (1970) 48.         
\item{24.}  L.D.~Faddeev and V.N.~Popov, {\smallit Phys. Lett.} 
{\smallbf 25B} (1967) 29. See also:
L.D.~Faddeev, {\smallit Theor. and Math. Phys.} {\smallbf 1}
 (1969) 3 (in Russian)
L.D.~Faddeev, {\smallit Theor. and Math. Phys.} {\smallbf 1} 
(1969) 1 (Engl. transl).
\item{25.}  S.~Mandelstam, {\smallit Phys. Rev.} {\smallbf 175} 
(1968) 1580, 1604. 
\item{26.}  G.~'t~Hooft, {\smallit Nucl. Phys.} {\smallbf B33} 
(1971) 173.
\item{27.}  C.~Becchi, A.~Rouet and R.~Stora, {\smallit Commun.
 Math.  Phys.} {\smallbf 42}  (1975)  127; 
     {\smallit id., Ann. Phys.(N.Y.)} {\smallbf 98} (1976) 287.
 See also:
     I.V.~Tyutin, Lebedev Prepr. FIAN39 (1975), unpubl.; R.~ Stora, 
     Carg\`ese lectures 1976; J.Thieri-Mieg, {\smallit J.Math.Phys.}
 {\smallbf 21} (1980)  2834; 
     L.~Beaulieu, and J.~Thieri-Mieg, {\smallit Nucl. Phys.} 
{\smallbf B197} (1982) 477.
\item{28.} S.L.~Adler, {\smallit Phys. Rev.} {\smallbf 177}
 (1969) 2426; see also  H.~Fukuda and 
Y.~Miyamoto, {\smallit Progr. Theor. Phys.} {\smallbf 4} (1949) 347.
\item{29.} J.S.~Bell and R.~Jackiw, {\smallit Nuovo Cim.}
 {\smallbf A60} (1969) 47.
\item{30.}    A.~Slavnov, {\smallit Theor. Math. Phys.} 
{\smallbf 10} (1972) 153 (in  Russian),  {\smallit Theor. 
   Math. Phys.} {\smallbf 10} (1972) 99 (Engl. Transl.)
\item{31.} J.C.~Taylor, {\smallit Nucl. Phys.} {\smallbf B33}
 (1971) 436.
\item{32.} G.~'t~Hooft, {\smallit Nucl. Phys.} {\smallbf B35}
 (1971) 167.
\item{33.} G.~'t~Hooft and M.~Veltman, {\smallit Nucl. Phys.}
{\smallbf B44} (1972) 189.
\item{34.} M.~Veltman, invited talk given at the Third International 
Symposium on the
History of Particle Physics, SLAC, June 24-27, 1992, section 15.
\item{35.} K.G.~Wilson, {\smallit Phys. Rev.}  {\smallbf D3}  (1971)  
1818;  K.G.~Wilson 
 and  M.E.~Fisher, {\smallit Phys. Rev. Lett.} {\smallbf 28} (1972) 240.
\item{36.} C.~Bollini and J.~Giambiagi, {\smallit Nuovo Cim.}
 {\smallbf 12B} (1972) 20.
\item{37.} J.~Ashmore,  {\smallit Nuovo Cim. Lett} {\smallbf 4}
 (1972) 289.
\item{38.} S.L.~Adler and W.A.~Bardeen, {\smallit Phys. Rev.} 
{\smallbf 182}  (1969)  1517;
   W.A.~Bardeen, {\smallit Phys. Rev.} {\smallbf 184} (1969) 1848;
 D.G.~Boulware, 
{\smallit Phys. Rev.}  {\smallbf D11} 
     (1975) 1404; {\smallit Phys. Rev.} {\smallbf D13} (1976) 2169.
\item{39.} S.~Weinberg, {\smallit Phys. Rev. Let.} {\smallbf 19} 
(1967) 1264.
\item{40.} A.~Salam  and J.C.~Ward, {\smallit Phys. Lett.} 
{\smallbf 13} (1964) 168;
 A.~Salam, in: Elementary Particle Theory, ed. N.~Svartholm
 (Stockholm, 1968).
\item{41.} Ch.~Llewellyn-Smith, {\smallit Phys. Lett.} {\smallbf B46} 
(1973) 233.
\item{42.} J.~Cornwall, D.~Levin and G.~Tiktopoulos, 
{\smallit Phys. Rev. Lett.} 
{\smallbf 30} (1973) 1268.
\item{43.} G.~'t~Hooft, announcement made at the Colloquium on
 Renormalization 
of Yang-Mills Fields, C.N.R.S., Marseille, June   19-23, 1972;
D.J.~Gross and F.~Wilczek, {\smallit Phys. Rev. Lett.} {\smallbf 
30} (1973) 1343;
H.D.~Politzer, {\smallit Phys. Rev. Lett.} {\smallbf 30} (1973)
 1346;
G.~'t~Hooft, {\smallit Nucl. Phys. \smallbf B61} (1973) 455; 
H.D.~Politzer, 
{\smallit Phys. Rep. 
\smallbf 14c} (1974) 129.
\item{44.} G.~'t~Hooft, {\smallit Nucl. Phys. \smallbf B79} 
(1974) 276;  
{\smallit Nucl. Phys. \smallbf B105} (1976) 538. 
\item{45.} A.M.~Polyakov, {\smallit JETP Lett. \smallbf 20} 
(1974) 194.
\item{46.} A.A.~Belavin, A.M.~Polyakov, A.S.~Schwartz and Yu.S.~Tyupkin, 
{\smallit Phys. 
     Lett. \smallbf 59B} (1975) 85.
\item{47.} G.~'t~Hooft, {\smallit Phys. Rev. Lett. \smallbf 37 }(1976) 8; 
     3432; R.~Jackiw and C.~Rebbi, {\smallit Phys. Rev. Lett.
 \smallbf 37} (1976) 172;  
C.G.~Callan Jr., R.F.~Dashen and D.J.~Gross, {\smallit Phys. Lett.
 \smallbf 63B} (1976) 334; 
     {\smallit Phys. Rev. \smallbf D17} (1978) 2717. 
\item{48.} G.~'t~Hooft, {\smallit Phys. Rev.} {\smallbf D14} 
(1976) 3432; Err. 
{\smallit Phys. Rev. \smallbf D18} (1978) 2199.
\item{49.} K.~Symanzik, {\smallit Nuovo Cim. Lett. \smallbf 6} 
(1973) 77
K.~Symanzik, {\smallit  Commun. Math. Phys. \smallbf 18} (1970) 227  
K.~Symanzik, {\smallit  Commun. Math. Phys. \smallbf 23} (1971) 49 
K.G.~Wilson, {\smallit Phys. Rev.\smallbf D3} (1971) 1818    
G.~'t~Hooft, {\smallit Nucl. Phys.\smallbf B62} (1973) 444;   
{\smallit Nucl. Phys. \smallbf B254} (1985) 11; 
{\smallit Phys. Lett.\smallbf 109B} (1982) 474; {\smallit ibid.
 \smallbf 119B} (1982) 369. 
\item{50.} G.~'t~Hooft and M.~Veltman, {\smallit Ann. Inst.
 Henri Poincar\'e, \smallbf 20} (1974) 69.
\item{51.} S.~Mandelstam, {\smallit Phys. Rep. \smallbf 23} (1976) 237;
H.J.~Lipkin, {\smallit Phys. Lett. \smallbf 45B} (1973) 267;
M.~Fritsch,  M.~Gell-Mann
     and H.~Leutwyler, {\smallit Phys. Lett. \smallbf 47B} (1973) 365;
     C.G.~Callan, R.~Dashen and D.~Gross, {\smallit Phys. Lett.
     \smallbf 66B} (1977) 375; C.G.~Callan, R.~Dashen and D.~Gross,
     {\smallit Phys. Lett. \smallbf 78B} (1978) 307;  A.M.~Polyakov,
     {\smallit Nucl. Phys.  \smallbf B120} (1977) 429; G.~'t~Hooft,
     {\smallit Phys. Scripta \smallbf 24}  (1981)  841; {\smallit
     Nucl.  Phys. \smallbf B190} (1981) 455.
\item{52.}  K.~Osterwalder and R.~Schrader, {\smallit  Commun. 
Math. Phys. \smallbf 31}  (1973)  83;
     {\smallit ibid. \smallbf 42} (1975) 281.
\item{53.} T.~Appelquist and J.~Carazzone, {\smallit Phys. Rev. 
\smallbf D11} (1975) 2856.
\item{54.} G.~'t~Hooft, {\smallit Nucl. Phys. \smallbf B72} (1974) 461; 
G.~'t~Hooft, {\smallit Nucl. Phys. \smallbf B75} (1974) 461  
{\smallit  Commun. Math. Phys. \smallbf 86} (1982) 449; {\smallit 
ibid. \smallbf 88} (1983) 1.
\item{55.} G.~'t~Hooft, in  ``Recent Developments in Gauge  Theories", 
 Carg\`ese 
     1979, ed. G.~'t~Hooft et al., Plenum Press, New York, 1980,  Lecture 
     III, reprinted in: ``Dynamical Symmetry Breaking,  a  Collection  of 
     reprints", ed.~A.~Fahri  et  al.,  World  Scientific,  Singapore, 
     Cambridge, 1982, p. 345.
\item{56.}  A.~Ringwald, {\smallit Nucl. Phys. \smallbf B330} (1990) 1; 
L.~McLerran,  A.~Vainshtein 
     and M.~Voloshin, {\smallit Phys. Rev. \smallbf D42} (1990) 171.
\item{57.}  V.~Rubakov, {\smallit JETP Lett \smallbf 33} (1981) 644; 
{\smallit Nucl. Phys. \smallbf B 203} (1982)  311;
     C.G.~Callan, {\smallit Phys. Rev. \smallbf D25} (1982) 2141,
     {\smallbf D26 } (1982)  2058;  {\smallit Nucl.  Phys. \smallbf
     B212} (1983) 391.
\item{58.} G.~'t~Hooft, in "The Whys  of  Subnuclear  Physics", 
 ed.  A.~Zichichi, 
     Plenum, New York/London, p. 943.

\end